# Impact of geometrical parameters on optical properties of negative curvature hollow-core fibers

G.K. Alagashev, A. D. Pryamikov[*], A.F. Kosolapov, A. N. Kolyadin, A.Yu. Lukovkin, and A.S .Biriukov

*Fiber Optics Research Center of Russian Academy of Sciences, 38 Vavilov street, Moscow, 119333, Russia*
[*]*pryamikov@fo.gpi.ru*

**Abstract:** We analyze the impact of geometrical parameters on such important optical characteristics of negative curvature hollow-core fibers (NCHCFs) as waveguide dispersion, waveguide losses and the structure of transmission bands. We consider both theoretically and experimentally the resonance effects and formation of band edges under bending in NCHCFs.

## 1. Introduction

Hollow-core microstructured optical fibers (HC MOFs) are waveguides which allow light to be localized in the hollow core filled with gases or liquids. High power radiation can be transmitted in NCHCFs due to a high level of light localization and low nonlinearity of the gas medium in the hollow core. This power level of transmitted radiation cannot be achieved in the solid core glass fibers [1].

The mechanisms of light localization in HC MOFs are different for different types of microstructured claddings. The first HC MOFs proposed in [2, 3] were hollow-core Bragg fibers with a multilayered cladding which can be described as a one dimensional photonic crystal with refractive index modulation in the radial direction. The light localization in hollow-core Bragg fibers is achieved by constructive interference which occurs under light scattering from the 1D photonic crystal cladding. The spectral ranges where radiation localized in the hollow core cannot propagate into the radial direction are called photonic band gaps.

By late XX century, a new type of HC MOFs with a 2D photonic crystal cladding was proposed [4]. These HC MOFs were called hollow-core photonic crystal cladding fibers (HC PCFs). The mechanism of light localization in HC PCFs is analogous to that in hollow-core Bragg fibers. It is also based on the presence of photonic band gaps of a 2D photonic crystal in the cladding. The cross section of the core cladding boundary in such fibers has a circular or polygonal form. The band edges in the case of hollow-core Bragg fibers are determined by the resonances – antiresonances in the individual layers of the cladding according to the ARROW model [5]. In contrast to the hollow core Bragg fibers HC PCFs have the complex topological structure of the silica air photonic crystal cladding and the photonic band gap is formed by three cladding resonators, namely, interstitial silica apexes, silica struts and air holes [6]. The periodical arrangement of these inclusions in the 2D photonic crystal cladding causes formation of different types of Bloch waves sets.

The third type of HC MOF known to date is hollow-core fibers with a Kagome lattice cladding [7]. The mechanism of light localization in such HC MOFs is different from that for HC PCFs and hollow-core Bragg fibers. In the literature this approach to the localization phenomenon is called the Inhibited coupling model (ICM) [7]. It has been demonstrated that this mechanism is akin to Von Neumann – Wigner bound states within a continuum [8]. Comparatively small losses in such HC MOFs are caused by a strong transverse – field mismatch between the core and cladding modes [7]. The authors of [9, 10] reported on the fabrication of the Kagome lattice hollow core photonic crystal fiber with a hypocycloid shaped core structure. They demonstrated based on the inhibited coupling model that the design of this core shape enhances the coupling inhibition between the core and cladding modes. In [11] the authors reported on the record transportation of milli - Joule energy pulses of 600 fs duration operating around 1030 nm in the hypocycloid – core Kagome HC PCF.

In recent years, the so-called negative curvature hollow core fibers (NCHCFs) without a photonic crystal cladding have been proposed and extensively investigated [12 – 14]. It was demonstrated to be possible to guide light in NCHCFs made of silica glass up to a wavelength of 8 µm despite a comparatively simple construction of the non-photonic crystal cladding [15]. According to our estimations NCHCFs with one row of capillaries in the cladding provide the highest degree of light power localization in air - approximately 99.993% [16]. This fact shows that the physical mechanism of light localization in NCHCFs is different from those in other types of HC MOFs. In our future papers, we will analyze this mechanism based on the ideas outlined in [17].

In this paper, we analyze the process of the band edge formation in NCHCFs with different cladding structures. In [18] the authors stated that it was not enough to consider the transmission attenuation properties of NCHCFs only in a qualitative way, for example, by a spatial power overlap between the core modes and cladding glass. In this work we propose a theoretical model based on the coupled mode theory to demonstrate a difference in mechanisms of band edge formation for short wavelength edges and long wavelength edges in NCHCFs with a certain type of rotational symmetry of the core – cladding boundary. In our opinion, this difference is due to the discreteness effect of the core – cladding boundary in NCHCFs on the process of band edge formation. The discreteness is determined by the periodical change in the curvature of the core-cladding boundary in the azimuthal direction and by the distances between neighboring capillaries in the cladding. Earlier, the effect of the curvature of the core cladding boundary on the loss level was investigated in [19, 20]. The difference between HC MOFs with different forms of the core-cladding boundary cannot be described by the ARROW model. The formation of the resonant structure of the long wavelength band edges can be caused by the excitation of nonlocal collective cladding modes. Further it will be shown that the intensity of such collective cladding modes and their spectral density depends strongly on the profile of the core cladding boundary in the azimuthal direction. This difference also leads to dispersion curves behavior not observed for other types of HC MOFs. As NCHCFs waveguide properties are dependent on the core cladding boundary profile, obtaining a few-mode waveguide regime in NCHCFs is also considered.

In this paper, we also provide an analysis of optical properties of NCHCFs under bending. In [13] we demonstrated for the first time that NCHCFs bend losses have resonant behavior dependent on the bend radius. The resonant increase in the bend loss occurs at certain values of the bend radius and wavelength. In this case, a resonant coupling occurs between the air core mode of NCHCFs and the Airy mode of the cladding capillary. This effect was confirmed theoretically in works [20, 21, 22]. In our paper we propose a theoretical model (numerical simulation) of the excitation process of the Airy modes. We also demonstrate that the bending leads to a transformation of the dispersion curves. This effect allows to manage the slope of the dispersion curves by changing the value of the bend radius. The simulations are in close agreement with the experimental data.

The paper has 7 sections. In Section 2, we propose a theoretical model describing the long wavelength band edge formation in NCHCF with a certain type of rotational symmetry of the core – cladding boundary. The necessary phase matching condition for resonant coupling between the air core modes and cladding modes are derived. The consideration is carried out based on the coupled mode theory. In Section 3 the process of the band edge formation in NCHCFs with touching and non touching capillaries in the cladding is considered. The difference between the mechanisms of short wavelength and long wavelength band edge formation is demonstrated. In Section 4, we discuss a possibility of obtaining a few mode regime in NCHCFs with different cladding structures. In the second part of the paper, we consider the system of transmission bands occurring under bending and corresponding transformation of dispersion curves. The process of Airy mode excitation in the cladding capillaries is investigated theoretically and experimentally. The conclusions are given in the last section.

## 2. Two phase matching conditions for optical momentums of the air core and cladding modes in waveguides with a certain type of rotational symmetry of the core – cladding boundary

In this section we consider a model describing coupling between the core and cladding modes in waveguides with a certain type of rotational symmetry of the core – cladding boundary. The analysis is based on the coupled mode theory [23]. For the first time, the resonant coupling between the air core modes and the cladding modes of hollow core polygonal tube fibers was demonstrated in [24]. These resonances occurred due to phase matching condition for azimuthal components of their optical momentums and had Fano profiles. Similar phase matching conditions occur in the case of excitation of whispering gallery modes in silica spherical micro cavities [25]. In our work [17] we showed that azimuthal mode indices $m$ of the scattered fields occurring under plane wave incidence on an individual dielectric cylinder are physically meaningful. They can be considered as a parameter describing the azimuthal component of the scattered fields optical momentums. For example, the plane wave scattering from a 1D metallic diffraction grating leads to an onset of resonances in the scattered field spectrum called Wood anomalies [26]. These anomalies have Fano and Lorentzian profiles and occur due to a resonant interaction between diffraction orders of the scattered field with different linear components of optical momentums. These linear components of optical momentums are determined by the values of the reciprocal lattice vector of the metallic diffraction grating. The same phenomenon can be observed under the plane wave scattering from an individual dielectric cylinder but in this case the anomalies occur due to a resonant interaction between diffraction orders of the scattered field with different azimuthal mode numbers.

It is known that the phase matching condition is a conservation law for linear components of optical momentums, for example, for propagation constants of the interacting waves [23]. In the case of waveguides with a certain type of rotational symmetry of the core cladding boundary it is possible to show that the coupling between the air core modes and the cladding modes occurs due to an additional phase matching condition. This condition is fulfilled for azimuthal mode numbers or, saying in other words, for azimuthal components of optical momentum of the core and cladding modes. Let us consider some general model of light propagation in the hollow core waveguide with a core – cladding boundary having a periodical modulation of dielectric susceptibility in the azimuthal direction with a period of

$\delta\phi = 2\pi/N$, where N is an integer. The dielectric susceptibility of the cladding can be cast in the form:

$$\varepsilon = \varepsilon^0 + \Delta\varepsilon(r,\phi),$$

where

$$\Delta\varepsilon(r,\phi) = \sum_{p=-\infty}^{+\infty} \varepsilon_p(r)e^{ipN\phi}, \quad (1)$$

where $p = \pm 1, 2, \ldots$. In this case, the wave equation can be cast as:

$$\left(\nabla^2 + \omega^2/c^2\left[\varepsilon^0 + \Delta\varepsilon(r,\phi)\right]\right)\vec{G} = 0, \quad (2)$$

where $\vec{G}$ denotes magnetic or electric fields in the complex representation:

$$\vec{G}(r,\phi,z,t) = \vec{A}(r,\phi)e^{i(\beta t - \omega t)}. \quad (3)$$

In (3) $\beta$ is a propagation constant of the mode and $\omega = kc$ is an angular frequency, where $k$ is a wave number. It is known that the complex representation of magnetic and electric fields in waveguides can be characterized by specifying axial components of the fields Ez and Hz. Next, we assume that the field amplitudes vary slowly in the azimuthal direction and the disturbance of the dielectric susceptibility (1) is weak. In this case, it is possible to apply the slow changing amplitudes method for solving (2) and the axial components of electric and magnetic fields in the cladding can be represented as:

$$E_z = \sum_{m=-\infty}^{+\infty} C_m(\phi)F_m(r)e^{im\phi}e^{i\beta_m z}, \quad (4)$$

$$H_z = \sum_{m=-\infty}^{+\infty} D_m(\phi)F_m(r)e^{im\phi}e^{i\beta_m z}.$$

Substituting (4) into (2) one obtains an equation for the coefficients Cm and Dm:

$$\frac{1}{r^2}\sum_{m=-\infty}^{+\infty}\left(\frac{\partial^2 C_m}{\partial\phi^2} - 2im\frac{\partial C_m}{\partial\phi}\right)F_m(r)e^{im\phi}e^{i\beta_m z} + \sum_{l=-\infty}^{+\infty}\Delta\varepsilon(r,\phi)C_l F_l(r)e^{il\phi}e^{i\beta_l z} = 0 \quad (5)$$

Here, it is assumed that function Fm(r) satisfies an equation:

$$\left(\frac{d^2}{dr^2} + \frac{1}{r}\frac{d}{dr} - \frac{m^2}{r^2} + k^2 n^2 - \beta_m^2\right)\sum_{m=-\infty}^{+\infty} C_m F_m(r)e^{im\phi} = 0.$$

According to the slow changing amplitudes method:

$$\frac{\partial^2 C_m}{\partial\phi^2} \ll m\frac{\partial C_m}{\partial\phi},$$

and taking into account (1) equation (5) can be represented as:

$$\frac{1}{r^2}\sum_{m=-\infty}^{+\infty} 2im\frac{\partial C_m}{\partial\phi}F_m(r)e^{im\phi}e^{i\beta_m z} = \sum_{l=-\infty}^{+\infty}\sum_{p=-\infty}^{+\infty}\varepsilon_p(r)C_l F_l(r)e^{i(l+Np)\phi}e^{i\beta_l z}. \quad (6)$$

Multiplying (6) by $F_m^*(r)e^{-im\phi}e^{-i\beta_m z}$ and integrating over the unit volume of the cladding one obtains:

$$C_m = \frac{1}{2imL}\sum_{l=-\infty}^{+\infty}\sum_{p=-\infty}^{+\infty} K_{lm}^p \int_0^{2\pi} C_l e^{i(l+Np-m)\phi}d\phi * \delta_{\beta_m \beta_l}, \qquad (7)$$

where $L = \int_{r_1}^{r_2} F_m(r)F_m^*(r)rdr$ and $K_{lm}^p = \int_{r_1}^{r_2} F_l(r)\varepsilon_p(r)F_m^*(r)r^3 dr$.

In (7) it is assumed that $e^{i(l+Np-m)j}$ is a fast oscillating function with respect to $C_l(\varphi)$. From (7) it can be seen that in addition to the well known phase matching condition for linear components of optical momentum of the interacting modes ($\beta_l \approx \beta_m$) there is also an additional phase matching condition for azimuthal components of the momentum $m = Np \pm l$. This expression coincides with the condition of effective coupling between the dielectric and the air core modes in the polygonal tube fibers [24]. The second type of a phase matching condition for low order air core modes occurs due to the azimuthal periodicity of the core – cladding boundary dielectric susceptibility (1). If $m$ and $l$ can be considered as numbers characterizing azimuthal components of optical momentums of the air core modes and cladding modes then $Np$ can be considered as an analogue of a reciprocal lattice vector in the case of a 1D metallic diffraction grating [26] or 1D photonic crystal [23]. As it was mentioned above, similar analogy was drawn in the case of excitation of Wood anomalies under the plane wave scattering from a single dielectric cylinder [17].

In the end of this section it is possible to draw certain conclusions regarding resonant coupling between the air core modes and cladding modes of the hollow core fibers with a certain type of rotational symmetry of the core – cladding boundary. If the waveguide structure has a discrete profile of the core – cladding boundary consisting of individual elements (as in the case of all solid band gap fibers [27]) the band edges are formed due to exciting electromagnetic states of these individual elements (the local cladding states). The coupling between the neighboring elements is weak. In this case, the first phase matching condition for the propagation constants of the air core modes and the cladding modes (linear components of optical momentums of the modes) play a decisive role in the process of band edge formation. If the waveguide structure has a continuous or quasi continuous profile of the core – cladding boundary with a certain type of rotational symmetry it is possible to excite cladding states which can be called collective cladding states. In this case, the coupling between the individual cladding elements is strong and the second phase matching condition can play an important role in the process of band edge formation. In the next section, we consider examples of excitation of the collective and local cladding states and their role in the formation of band edges using NCHCFs as an example.

## 3. Band edge formation and dispersion curve profiles in NCHCFs with respect to cladding geometry

In this section, we consider the behavior of transmission bands of NCHCFs with different geometrical parameters of the cladding and propose physical mechanisms of band edge formation. As it was mentioned above, long wavelength and short wavelength band edges in hollow-core Bragg fibers are determined by the ARROW model. These band edges are determined by the individual optical properties of the cladding elements, which are optical microresonators and have specific resonant frequencies. In the case of NCHCFs, the process of the band edge formation is more complicated and connected with an excitation of the collective cladding states with different types of rotational symmetry. As it was discussed above, the excitation occurs when corresponding phase matching conditions are fulfilled.

In the first samples of NCHCFs made of silica glass, the cladding capillaries touched each other, and the long wavelength band edges had resonant behavior in the mid IR spectral range [12]. These resonances could not be described by the ARROW model as the resonant wavelengths did not correspond to those of the cladding capillaries. Further, we proposed another design of the capillary arrangement in the cladding when the capillaries did not touch each other. This cladding construction allowed us to transmit light in an NCHCF made of silica glass up to a wavelength of 8 µm [15]. To determine the cause of the long wavelength resonances onset in the NCHCFs transmission bands we performed a number of simulations using commercial packet Femlab 3.1. The distance between the centers of two neighboring capillaries in the cladding (number of capillaries $N = 8$) was taken to be a changing parameter in the simulations. In the case of touching capillaries in the cladding the value of the parameter was $\Lambda = 2R_c$, where $R_c$ is the outer radius of the capillary. Several NCHCFs made of silica glass with 8 capillaries in the cladding were considered and the outer diameter of the capillary $d_{out}$ and the inner diameter $d_{in}$ were the same for all simulations. The core diameters of the NCHCFs were $D_{core} = 48, 72, 86$ µm, the thickness of the capillary wall was 3 µm, $R_c = 7$ µm and the distances between the centers of the capillaries were $\Lambda = 2R_c, 2.9R_c, 3R_c$. The process of the transmission bands transformation for different values of $\Lambda$ is shown in Fig. 1.

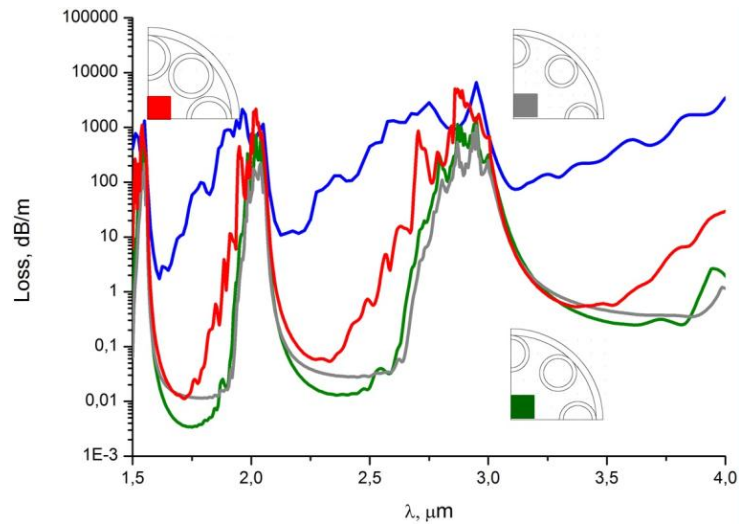

Fig. 1. The first three calculated transmission bands for NCHCFs made of silica glass with 8 capillaries in the cladding. The values of parameter $\Lambda = 2R_c$ (red line), $\Lambda = 2.9R_c$ (green line) and $\Lambda = 3R_c$ (grey line). The blue line corresponds to the transmission bands for the NCHCFs with 16 capillaries in the cladding. The other NCHCFs parameters are given in the text.

It can be seen from Fig.1 that the profiles of the transmission bands in the case of the NCHCFs with $\Lambda = 2R_c$ and $\Lambda = 2.9R_c, 3R_c$ are different. In the case of the NCHCFs with $\Lambda = 2R_c$ the long wavelength edges of the transmission bands have a resonant character. The resonances in the second and third transmission bands have a much higher Q-factor compared to that of the first (long wavelength) transmission band. It can also be seen that the resonances disappear when the value of parameter $\Lambda$ increases. It can then be concluded that the resonant behavior at the long wavelength band edges originated from the excitation of collective electromagnetic states of the cladding described in the previous section. This effect, in contrast to the ARROW model, has a nonlocal character, and this nonlocal resonant interaction between the air core modes and the cladding states leads to the reduction of an effective width of the transmission bands. The origin of the excitation of the collective cladding states is the coupling between the air core and cladding modes occurring under fulfilling the phase matching conditions for the axial and azimuthal components of the wavevectors. The simulations show that an increase in the distance $\Lambda > 3R_c$ in the case of NCHCFs with 8 capillaries in the cladding leads to an increase in the waveguide loss. In this case, the effect of loss reduction occurring due to an increase in the air core diameter is compensated by the loss increase occurring due to a decrease in the negative curvature of the core-cladding boundary.

The most striking instance of the excitation of the nonlocal cladding states can be observed for NCHCFs with 16 touching capillaries in the cladding (Fig. 1). The cladding capillaries had the same wall thickness as the NCHCFs with 8 capillaries in the cladding to eliminate the impact of the ARROW mechanism on the transmission spectrum. Moreover, the air core diameter was also the same as in the NCHCFs with 8 capillaries in the cladding with the value of parameter $\Lambda = 2R_c$. In this way, only the type of discrete rotational symmetry of the core – cladding boundary was changed. It can be seen from Fig. 1 that the resonances occurring at the long wavelength band edges bridge the long wavelength transmission bands almost completely, and the resonances have no such strong effect on the transmission bands only in the shortest wavelength transmission band (Fig .1) which leads to a waveguide loss reduction.

The positions of the short wavelength band edges for all the four NCHCFs considered (Fig.1) are essentially independent of $\Lambda$ parameter variations. It can be explained by the fact that the short wavelength band edges are determined by the resonant properties of the individual capillaries and corresponding local cladding states. In the case of NCHCFs with 16 capillaries in the cladding, the impact of the resonant properties of individual capillaries on the transmission bands formation is weaker than for NCHCFs with 8 capillaries in the cladding (Fig. 1 (red line)). We propose that for NCHCFs with 16 capillaries in the cladding the excitation of the collective electromagnetic cladding states in the long wavelength regions of the transmission bands occurs in a much broader spectral range due to an increase in the number of the excited collective states with higher azimuthal mode numbers $Np$. In other words, an increase in the number of the cladding capillaries leads to an increase in the excited collective cladding states with a period $\delta\phi = 2\pi p / N$ fulfilling the phase matching conditions not only at the long wavelength band edge of the transmission band but also at shorter wavelengths. For example, there are no resonant space harmonics with a sufficiently high value of the azimuthal mode number for NCHCFs with 8 capillaries in the cladding and with $\Lambda=2R_c$ at a wavelength of $\lambda = 3.3$ μm (Fig. 1). On the other hand, such harmonics can be excited for NCHCFs with 16 capillaries in the cladding and the profiles of the loss dependencies shown in Fig. 1 confirm our assumption.

To demonstrate the process of the collective states excitations at the long wavelength band edges we calculated overlap integrals for the first thirty air core modes in the second transmission band for NCHCFs with touching capillaries in the cladding (Fig. 1 (red)). The overlap integrals were calculated as:

$$I = \frac{\left|\int \vec{E}_{core}^* \vec{E}_{cladd} dS\right|^2}{\int \left|\vec{E}_{core}\right|^2 dS \int \left|\vec{E}_{cladd}\right|^2 dS} \tag{8}$$

where $\vec{E}_{core}$ and $\vec{E}_{cladd}$ are the complex electric fields of the air core and cladding modes. Then, we summarized their values to obtain the spectral dependence of the total overlap integral for all calculated cladding states (Fig. 2). As one can see from Fig. 2, the collective cladding states with different types of rotational symmetry are indeed excited at the long wavelength band edge of the transmission bands.

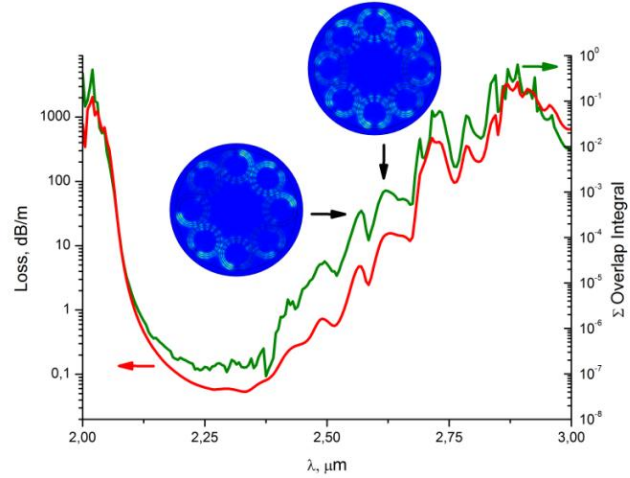

Fig. 2. The spectral dependence of the total overlap integral for the first thirty air core modes in the second transmission band. The excited collective cladding states are shown in the insets.

As a result, the transformation of the transmission bands occurring in the touching capillaries in the cladding leads to the transformation of dispersion curves. The dispersion curves were calculated for NCHCFs with 8 capillaries in the cladding (Fig. 1). It can be seen that the resonant behavior of the loss dependence at the long wavelength edges of the transmission bands for the NCHCFs with touching capillaries in the cladding leads to corresponding non monotonic behavior of the dispersion curves (Fig. 3). In terms of practical applications, the most interesting behavior of the dispersion curves is observed in the case of the NCHCF with $\Lambda = 3 \cdot R_c$ with a comparatively low transmission loss (Fig. 1).

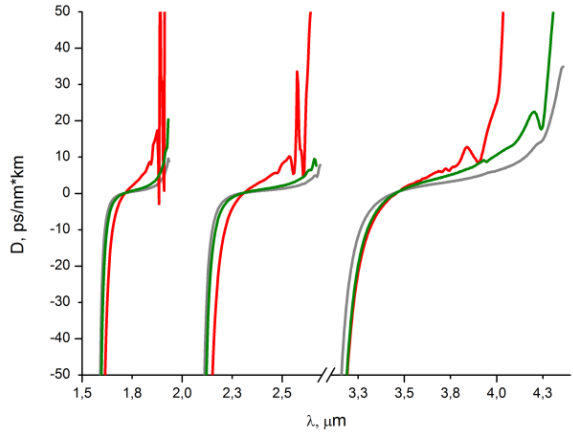

Fig. 3. The calculated dispersion curves for NCHCFs with 8 capillaries in the cladding corresponding to the transmission bands in Fig. 1.

In this case, both long and short wavelength band edges are formed by the ARROW mechanism without any additional resonances arising from the collective interaction of the cladding capillaries. This fact and strong light localization in the air core lead to the flat profile of the dispersion curves with a value of anomalous dispersion $D < 10$ ps/(nm*km) in all three transmission bands.

## 4. A few mode waveguide regime in NCHCFs

In this section, we analyze a possibility to realize a few mode regime in NCHCFs. Loss dependencies and spectral dependencies of $n_{eff}(\lambda)$ were calculated for NCHCFs with different values of $\Lambda$ for the first several air core modes in the second transmission band. The results are shown in Fig. 4 and 5.

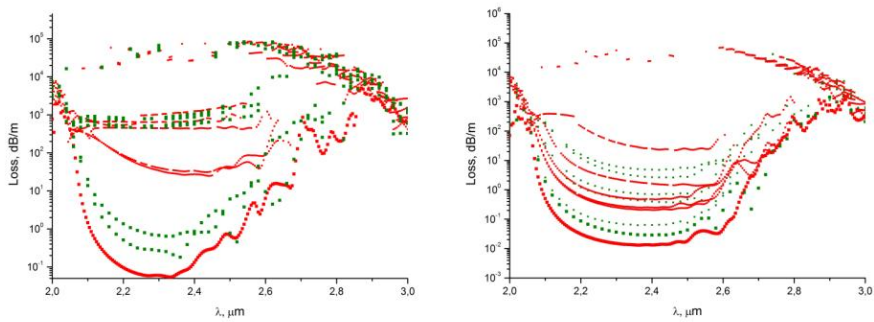

Fig. 4. (left) The loss dependence on the wavelength for the first 10 air core modes of NCHCFs with touching capillaries in the cladding in the second transmission band (Fig. 1); (right) analogous dependence for NCHCFs with non touching capillaries in the cladding ($\Lambda = 3R_c$).

Similar results can be obtained for other transmission bands. It can be seen that a few mode waveguide regime can be achieved for NCHCFs with touching capillaries in the cladding (Fig. 4(left)). As this takes place, $HE_{11}$ and $TE_{01}$ have the lowest loss. The higher order modes interact with the cladding much more intensively, which leads to the loss increase and strong coupling with the cladding states. Based on the results obtained in the previous section it is possible to assume that the higher order modes penetrate into the cladding much more deeply and interact with the collective cladding states at the long wavelength band edge (Fig. 4(left)).

At the same time, the spectral loss dependence for NCHCFs with non touching capillaries for the first 10 modes (Fig. 4(right)) differs significantly from the spectral loss dependence considered above. As a result, the coupling between the high order air core modes and the collective resonant states in the cladding is weakened significantly, with the high order modes having much lower loss in comparison with NCHCFs with touching capillaries in the cladding (Fig. 4(left)). This conclusion can be confirmed by considering the spectral dependencies of effective refractive indices of the air core modes (Fig. 5). In the case of NCHCFs with touching capillaries in the cladding (Fig. 5(left)) only two lowest order air core modes have close values of $n_{eff}(\lambda)$. The effective indices of the higher order modes differ from those for the lowest order modes most significantly in the neighborhood of the long wavelength band edge. At the same time, for NCHCFs with non touching cladding capillaries the values of $n_{eff}(\lambda)$ for a much higher number of the modes are close to each other.

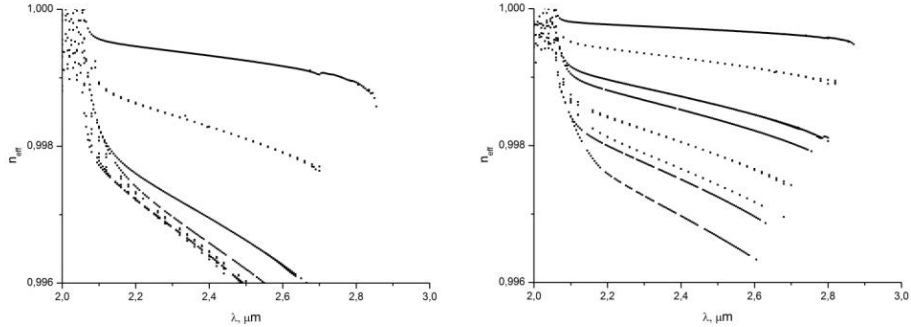

Fig. 5. (left) the spectral dependence of effective refractive indices of the air core modes for NCHCFs with touching cladding capillaries; (right) the spectral dependence of effective refractive indices of the air core modes for NCHCFs with non touching cladding capillaries

These results show that a few mode waveguide regime can be achieved in NCHCFs under the condition of a strong collective coupling between the cladding capillaries, which was described in the previous section.

## 5. Transformation of transmission bands under bending. Dispersion curves slope management.

In this section, we consider the impact of bending on the transformation of transmission bands and band edges. The geometrical parameters of the modeled NCHCFs corresponded to the geometrical parameters of the NCHCFs fabricated for the experiments. The parameters of the modeled NCHCFs equaled the averaged values of the fabricated fiber parameters. After that, we consider the impact of the bending on the slopes and profiles of the dispersion curves at all three values of $R_{bend}$.

To analyze the transformation of the transmission bands occurring under bending we plotted a diagram of loss dependence on the wavelength and the bend radius (Fig. 6). As one can see from Fig. 6, an additional system of band edges occurred at resonant values of the bend radius. As it was discussed above, the band edges at the resonant wavelengths occur due to the transverse resonances in the capillary wall or due to the excitation of the collective cladding mode. The band edges at the resonant values of the bend radius occur due to the transverse resonance of the capillary cavity. It can also be seen that the resonances locations in the long wavelength transmission bands shift to smaller values of the bend radius. It means that the excitation of the capillary cavity modes in the short wavelength regions occurs at smaller disturbances of the air core modes of NCHCFs under bending.

The excitation of the capillary cavity modes leads to several questions.

First of all, it is impossible to excite a leaky capillary cavity mode at the oblique incidence of the air core mode radiation when the NCHCF is straight. It can be assumed that the excitation of the capillary cavity mode occurs due to the refraction of the air core mode at the inner boundary of the capillary and penetration into its cavity under bending. An increase in the radiation intensity in the capillary wall depends on the bend radius value and the field structure in the wall. It is known that an increase in the effective refractive index along the curvature line plane (*x* axis) caused by the bending can be described by [28]:

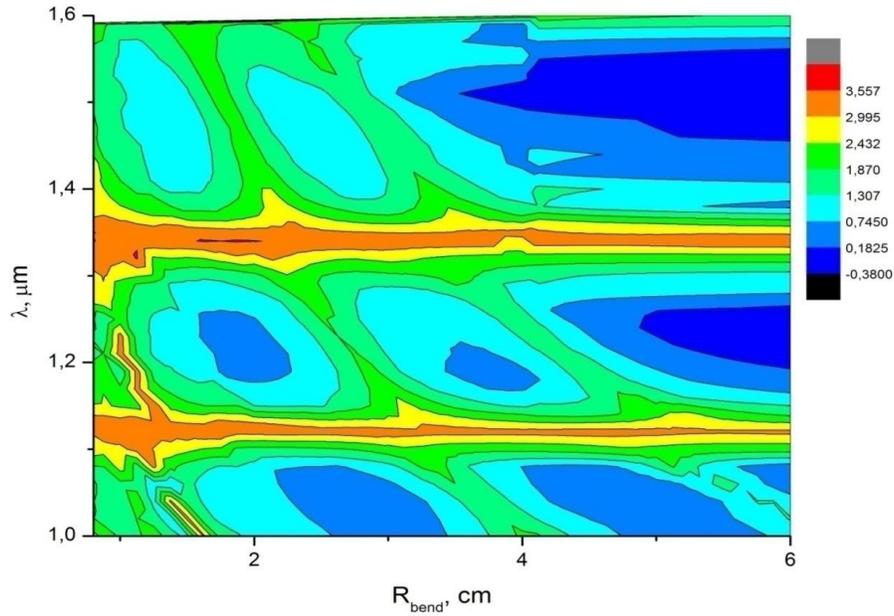

Fig. 6. A 2D diagram of the transmission bands occurring under change of the wavelength (the first three bands are shown) and under bending with the bend radius $R_{bend}$.

$$n(r,\phi) = n(1 + x/R_{bend}) \qquad (9)$$

Let us make some estimates taking into account that the air core radius of NCHCF $R_{core} \gg \lambda$ and the cladding capillary radius $R_{capill} \gg \lambda$, where $\lambda$ is a wavelength. Let the air core mode radiation fall onto the cladding capillary at an angle of $\theta = \cos^{-1}(n_{eff}/n) \approx 0.02$ rad., where $n_{eff} = 0.9998$ is an effective index of the fundamental air core mode at a wavelength of $\lambda = 1$ μm and a refractive index of the air core $n = 1$. Then, the angle of the refractive beam can be estimated by:

$$\sin^{-1}(\sin\theta_1/n_2) \approx 0.7608 \text{ rad.},$$

where $\theta_1 = \pi/2 - \theta$ is the incidence angle of the air core mode on the cladding capillary and $n_2$ is the refractive index of silica glass. Correspondingly, the total internal reflection angle in the case of light incidence on the boundary glass-air is approximately equal to 0.761 rad. It means that to a very close approximation the radiation falling on the cladding capillaries is reflected from the internal capillary wall in the regime of total internal reflection. Only a small part of the radiation penetrates into the capillary cavity. This condition is disturbed under bending and the increase in the effective refractive index (9). In this case, breaking the regime of total internal reflection can be observed. Therefore, the excitation of the capillary cavity mode occurs when the regime of total internal reflection turns into the refraction regime.

Let us make one more estimate. One can reasonably assume that the capillary cavity modes can be described by the following transverse resonance condition:

$$J_\nu\left(k_t^c a\right) = 0 \tag{10}$$

where $k_t^c$ is the transverse component of the wavevector in the capillary air core and $a$ is the inner radius of the capillary. The first roots of Bessel functions with $\nu = 0, 1$ equal $x_1 = 2.4$ and $x_2 = 3.8$, correspondingly. In the straight fiber at $a = 15$ μm and the effective index of the fundamental air core mode $n_{eff} = 0.9998$ the value of the argument in (10) doesn't exceeds $x_1$ and $x_2$. In this case, the excitation of the capillary cavity modes is impossible. When the bend radius decreases, for example, $R_{bend} = 5$ cm (Fig. 6) at the air core radius of the NCHCF $R = 30$ μm the argument in (10) is $k_t^c a > x_1$. It means that the condition of the transverse resonance (10) can be fulfilled at least for the fundamental air core mode. The higher order transverse resonances corresponding to higher order leaky modes of the capillary cavity occur with a further decrease in the bend radius. From the above, it is possible to assume that the decrease of the bend loss and suppression of excitation of the capillary cavity modes can be achieved by decreasing the capillary radius $a$. It can be achieved, for example, by increasing a number of the capillary in the cladding.

As one can see from Fig. 6, locations of the band edges caused by the bending in the long wavelength transmission bands are offset to smaller values of the bend radius. It means that the excitation of the capillary cavity modes occurs in the short wavelength transmission bands under smaller disturbances of the electromagnetic fields of the air core modes. Let us consider this process in greater detail.

Above it was observed that the resonant condition (10) cannot be fulfilled in the straight NCHCF because the value of $k_t^c a$ does not attain the root values of the Bessel functions. Under bending, the transverse component of the local wavevector at the inner side of the capillary wall can be represented as:

$$k_t^c = \frac{2\pi}{\lambda}\sqrt{n^2(1+x/R_{bend})^2 - n_{eff}^2} \tag{11}$$

where $\lambda$ is a wavelength. Here, we neglect the capillary wall thickness $h$ in comparison with the air core radius $R$ ($h<<R$). Assuming that $n^2 \approx n_{eff}^2$ in the capillary cavity one obtains an approximate value of the transverse component of the local wavevector:

$$k_t^c \approx \frac{2\pi n}{\lambda}\sqrt{\frac{2x}{R_{bend}}} \quad (12)$$

Substituting (12) into (10) one obtains the condition of the transverse resonance:

$$J_\nu(\frac{2\pi n}{\lambda}\sqrt{\frac{2R}{R_{bend}}}a) \approx 0 \quad (13)$$

It can be seen from (13) that the condition of the first transverse resonance is fulfilled under decreasing the bend radius first in the shortest wavelength transmission band (Fig. 6) and then in all further bands.

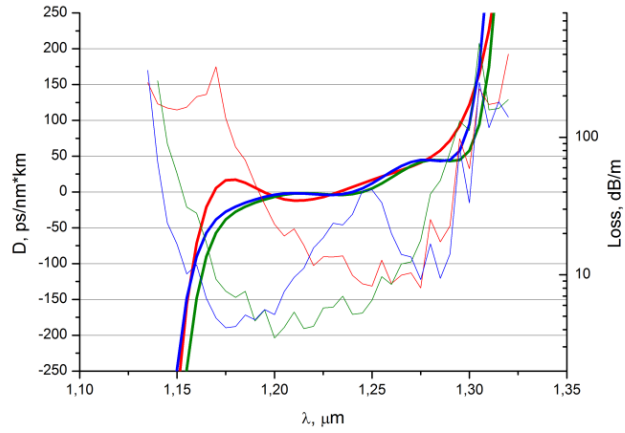

Fig. 7. Transformation of dispersion curves in the transmission band from 1.15 µm to 1.4 µm (Fig. 5) at $R_{bend}$ = 3 cm (red line), 3.5 cm (green line) and 4 cm (blue line).

Another interesting NCHCF property is the change of dispersion curve profiles of the air core modes which occurs in the bent fiber under excitation of the capillary cavity modes. For example, let us consider the behavior of the dispersion curves of the fundamental air core mode in the transmission band from 1.15 µm to 1.35 µm at the bend radius $R_{bend}$ = 3, 3.5 and 4 cm (Fig. 6). These three dispersion curves correspond to the bend radius in the vicinity of the resonance, at the slope of the resonance and away from the resonance (Fig. 7). It is also possible to manage a slope of the dispersion curves to a greater or a lesser extent depending on the parameter $\Lambda$. As in the processes described in Section 3, it is possible to assume that in the case of touching capillaries in the cladding there is a more dramatic change of the dispersion curves slope in the anomalous dispersion region due to the interaction between the NCHCF air core modes and the non local cladding states occurring at the long wavelength region (Fig. 1, 3). In other words, the interaction of the air core modes with the cladding under bending occurs much more effectively at the long wavelength band edges. It can be seen from

Fig. 7 that the zero dispersion wavelength also undergoes a shift and this dispersion curves behavior of NCHCFs under bending can be used for managing their dispersion characteristics.

**6. Experimental confirmation of the band loss resonant behavior in NCHCFs. Sequential excitation of the capillary cavity modes under bending: numerical simulations and experimental data.**

In this Section we investigate experimentally a set of the transmission bands occurring in NCHCFs under bending. Then we demonstrate experimentally a sequential excitation of the capillary cavity modes. In [13] it was theoretically shown that a coupling between the fundamental air core mode and the capillary cavity modes was possible and this coupling is manifested in the resonant dependence of the optical loss on the bend radius. Later, the resonant coupling between the modes was mentioned in [20 – 22]. However, up to this time the capillary cavity modes excited under bending have not been observed experimentally. In this work we managed to observe the modes experimentally. To this end, we used an NCHCF with 8 touching capillaries in the cladding. The NCHCF had the outer diameter of 125 µm, the core diameter of 40 µm, the inner capillary diameter of 27 µm and wall thickness of 3.5 µm. The experimental setup is shown in Fig. 8.

To measure the distribution of light intensity at the fiber end face a silicon camera with a microscope objective was used (Fig. 8(upper)). A Ytterbium fiber laser with a generation wavelength of 1.064 µm was used as a light source. To observe the capillary cavity modes at the fiber end face it was necessary that the straight part of the fiber just after the bend be as short as possible due to a strong attenuation of the capillary cavity modes. For this reason the fiber was held in the immediate vicinity of the end face and the bend part of the fiber was located just beside the fixation point. The fiber was bent at an angle of $90^0$ and the bend radius was controlled by micro adjustments.

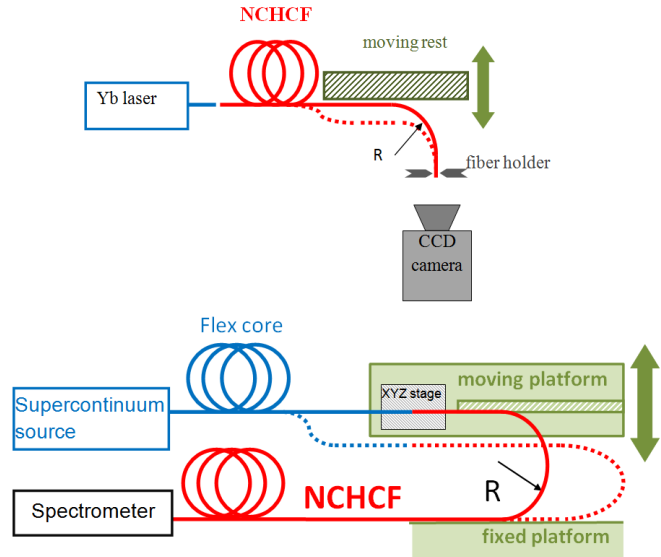

Fig. 8. (upper) Experimental setup for observing the distribution of light intensity in the near field; (bottom) experimental setup for measuring the transmission spectrum of the fiber at different values of the bend radius.

Another experimental design was used for measuring a dependence of the transmitted signal intensity on the bend radius value (Fig. 8(bottom)). A supercontinuum source (Fianium) was used as a light source. The spectrum of the transmitted radiation was measured by the spectrum analyzer Ando AQ6317B. We put an emphasis on the fiber being bent only at one section while the other fiber sections were undisturbed. To this end, the joint place between the NCHCF and the delivery fiber was put on a moving platform.

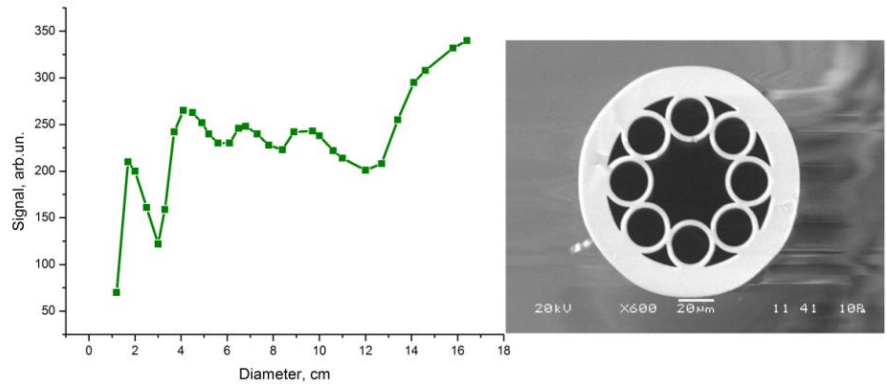

Fig. 9. (left) dependence of transmission light intensity on the bend diameter at a wavelength of 1.06 μm; (right) SEM photograph of the fiber cross section.

As one can see from Fig. 9, there are several minimums on the signal versus bend diameter curve. Each corresponds to the effective refractive index of the fundamental air core mode being equal to the effective refractive index of a cladding mode.

Comparing the diagram on Fig. 6 for the ideal NCHCF and the analogous diagram for the real one with the same geometry parameters (Fig. 10(left)) it is possible to draw the conclusion that these parameters have strong impact on the NCHCF optical properties. In the case of a real NCHCF all sizes and thickness of the capillary walls vary during the drawing process and differ from each other. This fact leads to blurring the distribution of the resonances in the diagram (Fig. 10(left)).

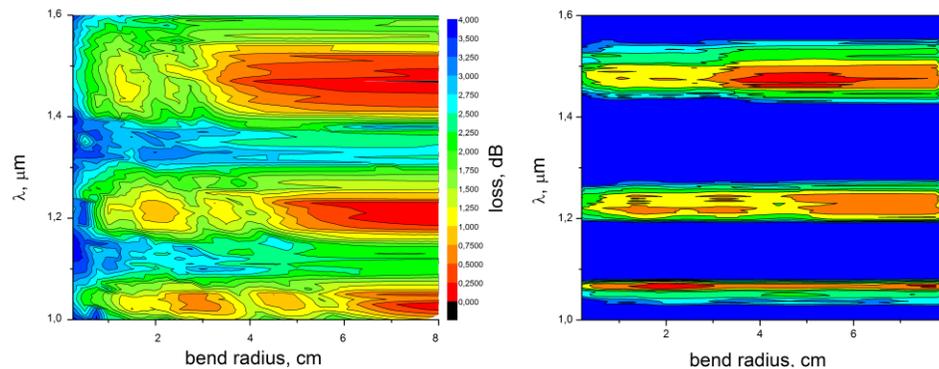

Fig. 10. (left) the calculated transmission bands for the real NCHCF; (right) dependence of transmitted light intensity on the value of the bend radius under excitation by the supercontinuum source.

The diagram shown in Fig. 10 (left) is in close agreement with an analogous diagram for the ideal NCHCF shown in Fig. 6. The experimental data is also in close agreement with numerical data (Fig.10 (right)). In particular, two resonant minimums are observed in the transmission band at the bend radius of 5 cm and in the vicinity of 3 cm.

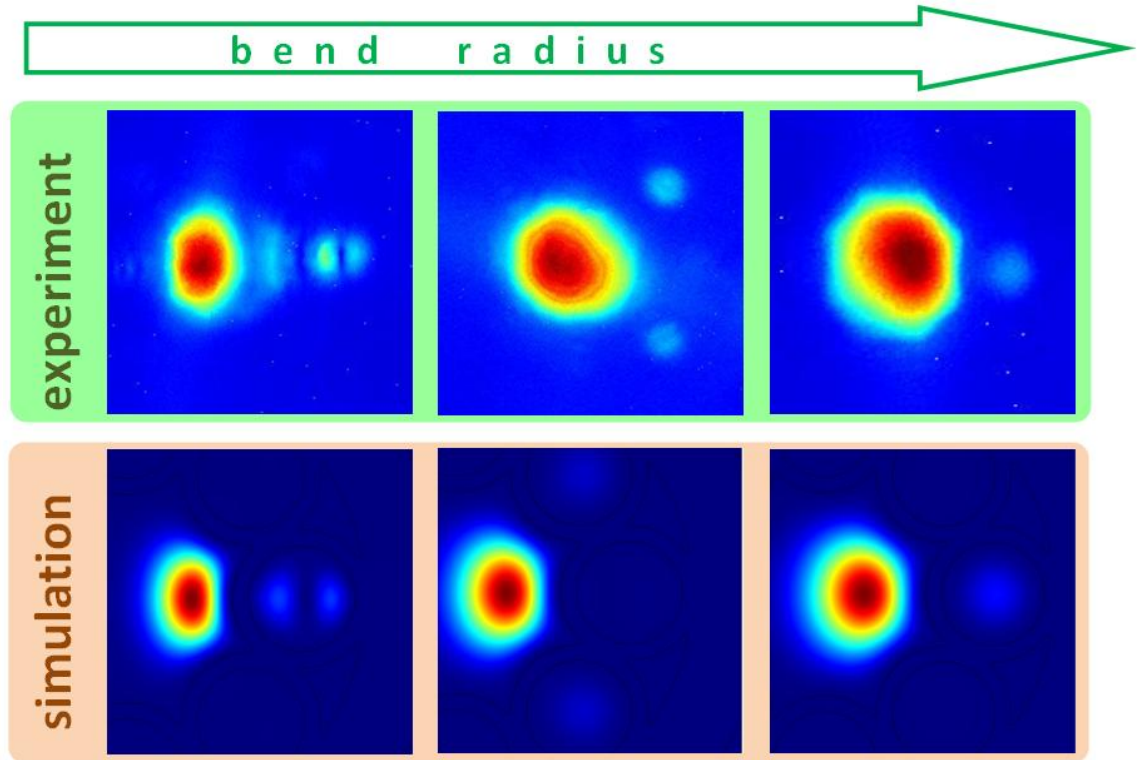

Fig. 11. Experimental and calculated near fields distributions of the air core modes corresponding to the first three resonances in the capillary cavity. The bend radii for the calculated near fields distributions are $R_{bend}$ = 1.6, 3.6 and 5.6 cm at wavelength of $\lambda$ = 1.06 μm. The bend axis is on the left side.

In Fig. 11 shows the near field distributions of the air core modes measured experimentally and numerically calculated. These near field distributions correspond to the first three resonances occurring under bending. The $HE_{11}$ mode is excited in the capillary lying in the bending plane at the bend diameter about 12 cm.

As the bend diameter decreases the fundamental mode in the considered capillary vanishes and two fundamental modes are excited in the neighboring capillaries. With the further decrease in the bend diameter, $TE_{01}$ is excited again in the capillary with the maximum bend radius.

We can clarify the causes of this selective excitation of the individual capillaries under bending by studying transverse resonance (10). Let us consider bending in a cylindrical coordinate system with $z$ – axis coinciding with the NCHCF axis. Let the bend be applied to the fiber into the capillary direction with the center of ($\rho$, 0) in the cylindrical coordinate system. Then, the expression for the effective refractive index (9) can be rewritten as:

$$n(r,\phi) = n(1 + \rho \cos\phi / R_{bend}) \tag{14}$$

where $\rho$ is the radius vector drawn from the fiber center to the capillary wall and $\varphi$ is a counterclockwise azimuthal angle. It can be seen from (14) that the resonant condition (10) for different values of the azimuthal angles $\varphi$ can be satisfied by the local transverse component of the air core modes wavevector $k_t^c(\rho,\varphi)$ (10, 11) at a given value of the bend radius. Thus, the maximum increase of the local effective refractive index and, correspondingly, the refraction and transmission of the air core mode intensity to the capillary cavity mode occurs in the direction of $\varphi = 0$. As the argument in (10) achieves its first resonant value the fundamental capillary cavity mode in the capillary with the center at ($\rho$, 0) is excited. According to (14), with the further bend radius decrease the local value of the propagation constant is changed. This change leads to a mismatch for the condition of the transverse resonance (10) in the direction of ($\rho$, 0) and to the excitation of the fundamental capillary cavity modes in the capillaries with centers at ($\rho cos(\pi/8)$, $\rho sin(\pi/8)$), ($\rho cos(15\pi/8)$ and $\rho sin(15\pi/8)$). With the further bend radius decrease this process is repeated for the second capillary cavity mode (Fig. 11).

## 7. Conclusions

In this work the dependence of NCHCF optical properties on geometrical parameters was determined. In particular, the process of band edges formation was analyzed. It was shown that the formation of long wavelength band edges strongly depends on the distance between the centers of neighboring cladding capillaries and cannot always be described in terms of the ARROW model. Such resonant behavior occurs due to excitation of the collective cladding states (i.e. electromagnetic states of the whole cladding). Short wavelength band edges occur due to excitation of the local cladding states (i.e. electromagnetic states of the individual cladding capillary). The formation of a few mode waveguide regime in NCHCFs also largely depends on the distance between the centers of the neighboring cladding capillaries. The analyses of optical properties of NCHCFs occurring under bending demonstrated the presence of new type of transmission bands. In this case, the band edges occur due to resonant properties of the capillary cavities (leaky modes of the cavities). A model describing the excitation of the capillary cavity modes was proposed. A possibility of dispersion curves transformations in normal and anomalous dispersion regions under bending was demonstrated. The experimental data is in close agreement with the theoretical results. The investigated physical processes occurring under change of the NCHCF geometrical parameters allow to optimize the NCHCF structure in different spectral ranges for various practical applications.